\documentclass[
review,number,sort&compress
]{elsarticle}
\usepackage{amsmath}
\usepackage{amssymb}
\usepackage{paralist}
\usepackage{hyperref}%
\usepackage{graphicx}% Include figure files
\usepackage{dcolumn}% Align table columns on decimal point
\usepackage{bm}% bold math
\usepackage{soul}
\usepackage{color}

\usepackage[mathlines]{lineno}% Enable numbering of text and display math
\definecolor{darkblue}{rgb}{0.0,0.0,0.66}
\hypersetup{pdfstartview={XYZ null null 1.5},colorlinks,breaklinks,
            linkcolor=darkblue,urlcolor=darkblue,
            anchorcolor=darkblue,citecolor=darkblue}

\def\sf{spontaneous f\/ission}

\def\root{\textsc{root}}
\def\srim{\{SRIM\}}
\def\mg{$\mu$g/cm$^{2}$}

\journal{Nuclear Instruments and Methods}

\begin{document}

%\preprint{}
\begin{frontmatter}

\title{Development of position-sensitive time-of-flight spectrometer for fission fragment research}
\author[lanl]{C.W. Arnold\corref{cor}}
\cortext[cor]{Corresponding author.}
\ead{arnold@lanl.gov}
\author[lanl]{F. Tovesson}
\author[lanl]{K. Meierbachtol}
\author[lanl]{T. Bredeweg}
\author[lanl]{M. Jandel}
\author[lanl]{H. J. Jorgenson}
\author[lanl]{A. Laptev}
\author[lanl]{G. Rusev}
\author[lanl,csm]{D. W. Shields}
\author[lanl]{M. White}%
\author[unm]{R. E. Blakeley}
\author[unm]{D. M. Mader}
\author[unm]{A. A. Hecht}
\address[lanl]{Los Alamos National Laboratory, Los Alamos, NM 87545}
\address[csm]{Colorado School of Mines, Golden, CO 80401}
\address[unm]{University of New Mexico, Albuquerque, NM 87131}
\date{\today}% It is always \today, today,
             %  but any date may be explicitly specified

\begin{abstract}
A position-sensitive, high-resolution time-of-flight detector for fission fragments has been developed.  The SPectrometer for Ion DEterminiation in fission Research (SPIDER) is a $2E-2v$ spectrometer designed to measure the mass of light fission fragments to a single mass unit. The time pick-off detector pairs to be used in SPIDER have been tested with $\alpha$-particles from $^{229}$Th and its decay chain and $\alpha$-particles and spontaneous fission fragments from $^{252}$Cf. Each detector module is comprised of a thin electron conversion foil, electrostatic mirror, microchannel plates, and delay-line anodes. Particle trajectories on the order of 700 mm are determined accurately to within 0.7 mm. Flight times on the order of 70 ns were measured with 200 ps resolution FWHM.  Computed particle velocities are accurate to within 0.06 mm/ns corresponding to precision of 0.5\%. 
An ionization chamber capable of 400 keV energy resolution coupled with the velocity measurements described here will pave the way for modestly efficient measurements of light fission fragments with unit mass resolution. 

\end{abstract}

\begin{keyword}
Spectrometers \sep fission \sep fission product yields \sep TOF \sep $2E-2v$
%% keywords here, in the form: keyword \sep keyword

%% MSC codes here, in the form: \MSC code \sep code
%% or \MSC[2008] code \sep code (2000 is the default)

\end{keyword}

\end{frontmatter}                              %display desired
%\maketitle

\section{\label{sec:level1}Introduction\protect}

% How did detector come to be?
Mass spectrometers for fission fragment studies have been used for decades \cite{Starzecki82, Oed84, Wilken84, Schilling87, CORSET}. %come in several flavors and can be optimized for user-specific needs \cite{}.  
We present here the results of tests designed to calibrate the detectors which measure particle velocities for the SPectrometer for Ion DEtermination in fission Research (SPIDER).  SPIDER is designed to be a $2E-2v$ spectrometer that infers the masses of a pair of fission fragments by measuring their velocities and kinetic energies ($E_{k}$).  Particle velocity is determined by combining measurements of a particle's path length and time-of-flight (TOF). When also combined with $E_{k}$ measurements, the non-relativistic mass of the particle is uniquely defined as
\begin{equation}
m = 2\frac{E_{k}}{v^{2}}
\label{eq:M}.
\end{equation}
% What is required of the equiptment to produce the desired results?
To achieve the requisite precision of a single mass unit ($\delta m$ $\leq$ 1 amu) the resolution requirements for path length, TOF and $E_{k}$ are straightforwardly constrained by
\begin{equation}
\frac{\delta m}{m} = \sqrt{\left(\frac{\delta E}{E}\right)^{2} + \left(2\frac{\delta t}{t}\right)^{2} + \left(2\frac{\delta r}{r}\right)^{2}}
\label{eq:UNC}.
\end{equation}
With an achievable energy resolution of 400 keV for 100-MeV fragments \cite{Oed84}, path length accuracy of 0.7 mm over  700 mm, and TOF resolution of 250 ps (FWHM) for 50-ns transit times, SPIDER will meet the demands for unit mass resolution for light fission fragments. 

Two standard sources, $^{229}$Th and its decay chain, and $^{252}$Cf, were used to determine TOF and path length resolution of the present system.  The $\alpha$-particles from $^{229}$Th and its decay chain yield six well resolved $\alpha$-particle energies.   The $\alpha$-particles and spontaneous fission fragments from $^{252}$Cf test our system with heavy ions in addition to $\alpha$-particles of known energy.    

This manuscript will give detailed descriptions of the detector components and their performance in section \ref{sec:DD} followed by a discussion of the experiments performed and the data collected in section \ref{sec:ExperimentandDP}.  Section \ref{sec:sim} describes simulations of the experimental setup and how they compare with the data. Finally, results and implications are discussed in section \ref{sec:results}.

\section{\label{sec:DD}Detector Description\protect}
\begin{figure*}
\includegraphics[width=1\textwidth]{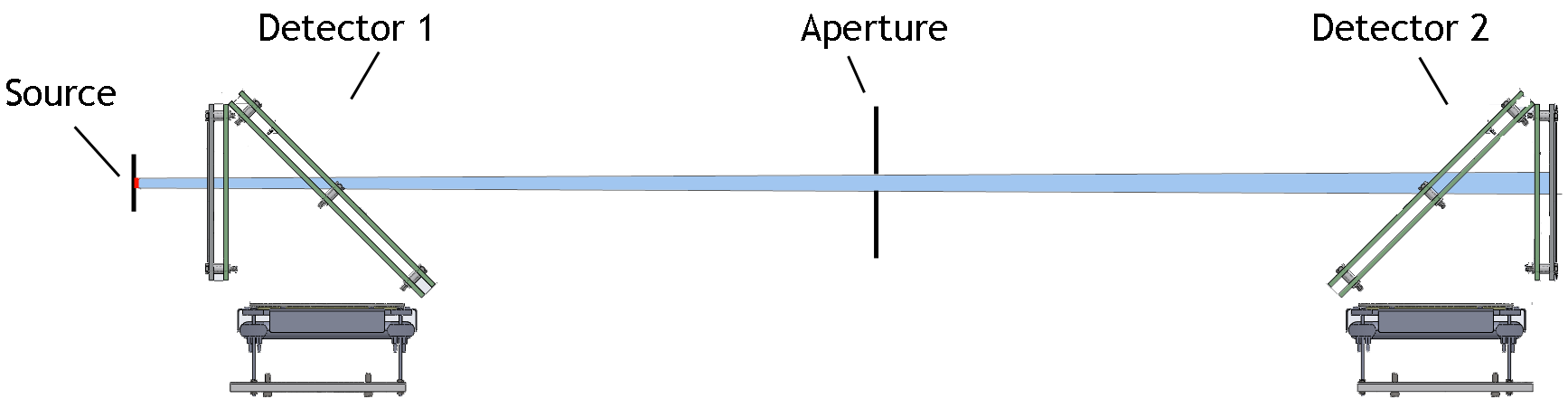}
\caption{\label{Fig:TOFdiagram}(Color Online) Schematic configuration of source and detector placement for TOF measurements in SPIDER.  The aperture restricts the trajectories (blue band) of the $\alpha$-particles which can be detected in the second detector. The aperture is removed for normal operation.}
\end{figure*}
The TOF spectrometer consists of an evacuated chamber with two time pick-off detectors separated by 50 - 70 cm. An $\alpha$-particle or a \sf~source is placed close to one of the time pick-off detectors. The source and the detectors are arranged such that emitted particles can pass though both detectors. % in order to measure their time-of-flight. 
The detectors were mounted to platforms that were attached to rails welded into the base of the chamber, allowing for changes to the distance between the detectors as desired.  A schematic drawing of the setup is given in Fig. \ref{Fig:TOFdiagram}.
 
The time pick-off detectors (see Fig.~\ref{FIG:MIR2pan}) are assemblies of electron conversion foils, electrostatic mirrors, microchannel plates (MCPs) and delay line anodes (DLAs). Both assemblies are kept under vacuum ($\leq$ 10$^{-6}$ Torr) during operation.
Herein we give details on the characteristics of each component and their purpose in the assembly.

\subsection{\label{sec:carbonfoils}Conversion Foils\protect}
As charged particles pass through a thin solid material they liberate electrons.  The typical energy of these secondary electrons (SE) is a few eV~\cite{thinfoils}. 
These SE can be accelerated under high voltage differences, and collected to give information about the location and the time that the SE were created.   
Carbon foils with thicknesses ranging from 20 to 100 \mg~and Mylar foils with evaporated gold layers were tested in the present setup with considerations made regarding straggling, efficiency and fragility. To minimize the effects of straggling, thin foils are ideal.  However, thinner foils are more prone to fail structurally. Mylar is stronger than carbon and the high $Z$ of gold increases MCP signal size and thus detection efficiency. However, the Mylar and gold foils induce too much straggling when considerations for future $E_{k}$ measurements are made. Thus, carbon foils of approximately 20$\mu$g/cm$^{2}$ were determined to be optimal for fission fragment measurements.   

\subsection{\label{sec:electronmirror}Electrostatic Mirror\protect}
\begin{figure}[t]
\includegraphics[width=0.45\textwidth]{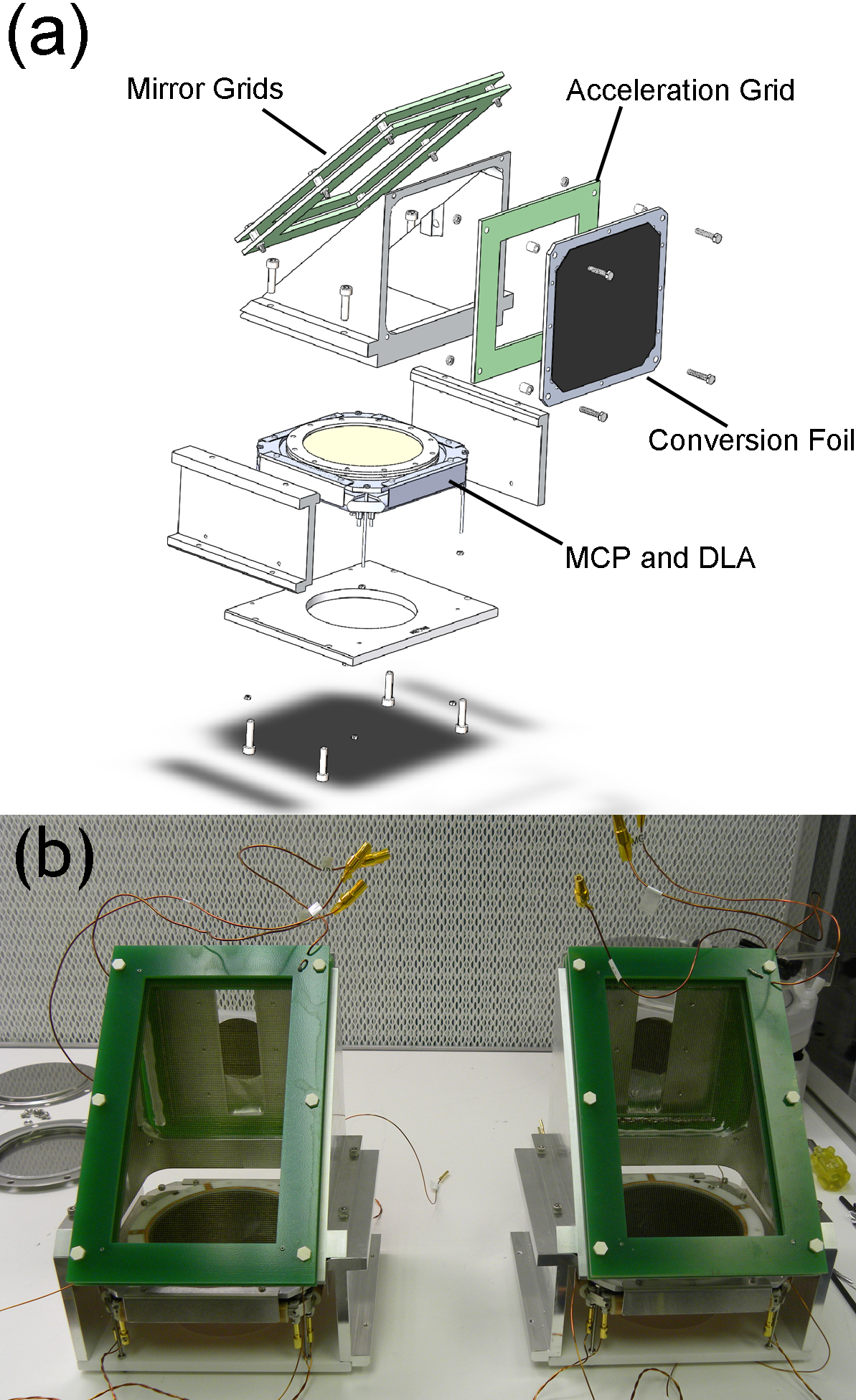}
\caption{\label{FIG:MIR2pan}(Color Online) (a) Exploded view of a time pick-off detector. (b) Two complete time pick-off detectors constructed for the present tests. }
\end{figure}

Electrostatic mirrors are designed to be mostly transparent to heavy ions while reflecting SE onto a device like a microchannel plate.  The first element in our electrostatic mirror, the conversion foil, is mounted parallel to a wire frame 1 cm away. The wire frames are constructed with 0.02-mm thick gold-plated tungsten wires spaced 1 mm apart. Each frame nominally has a geometrical transparency of 98\%. High negative voltage (-2000 to -3000 V) is applied to the foil while the wire frame is grounded to create a potential difference that accelerates SE.   After accelerating, the SE ideally drift through a field-free region to the mirror grids which are two wire frames (identical wire type and spacing) spaced 1 cm apart, and mounted at 45 degrees with respect to the conversion foil and first wire frame. The first mirror frame is grounded and the second mirror frame is at high negative voltage (-2000 to -3000 V). The wires on one mirror frame are orthogonal to the wires on the other mirror frame.  The frame with its wires stretched in the long (vertical) dimension has the same transparency as the first accelerating frame (98\%).  The frame with wires stretched along the short (horizontal) dimension has lower transparency (97\%) coresponding to a factor of $\sqrt{2}$ reduction in the space between wires that results from the angle of the grid. 
Figure \ref{FIG:MIR2pan}~shows an exploded drawing of our time pick-off detector as well as a photo of the completed version.
%These grids reflect electrons onto the next element in the system (the MCPs).  
A typical time from creation of SE to collection in the MCP is about 3 ns.  This SE transit time should be nearly identical for the $start$ and $stop$ detectors. Averaged differences in the transit time from two different electrostatic mirrors represent a determinable systematic factor in the TOF measurements that does not affect precision.  However, small variations in this transit time are a source of uncertainty in the measurement of TOF.

\subsection{\label{sec:MCPsandDLAs}MCPs and DLAs\protect}
\begin{figure}[t]
\includegraphics[width=0.45\textwidth]{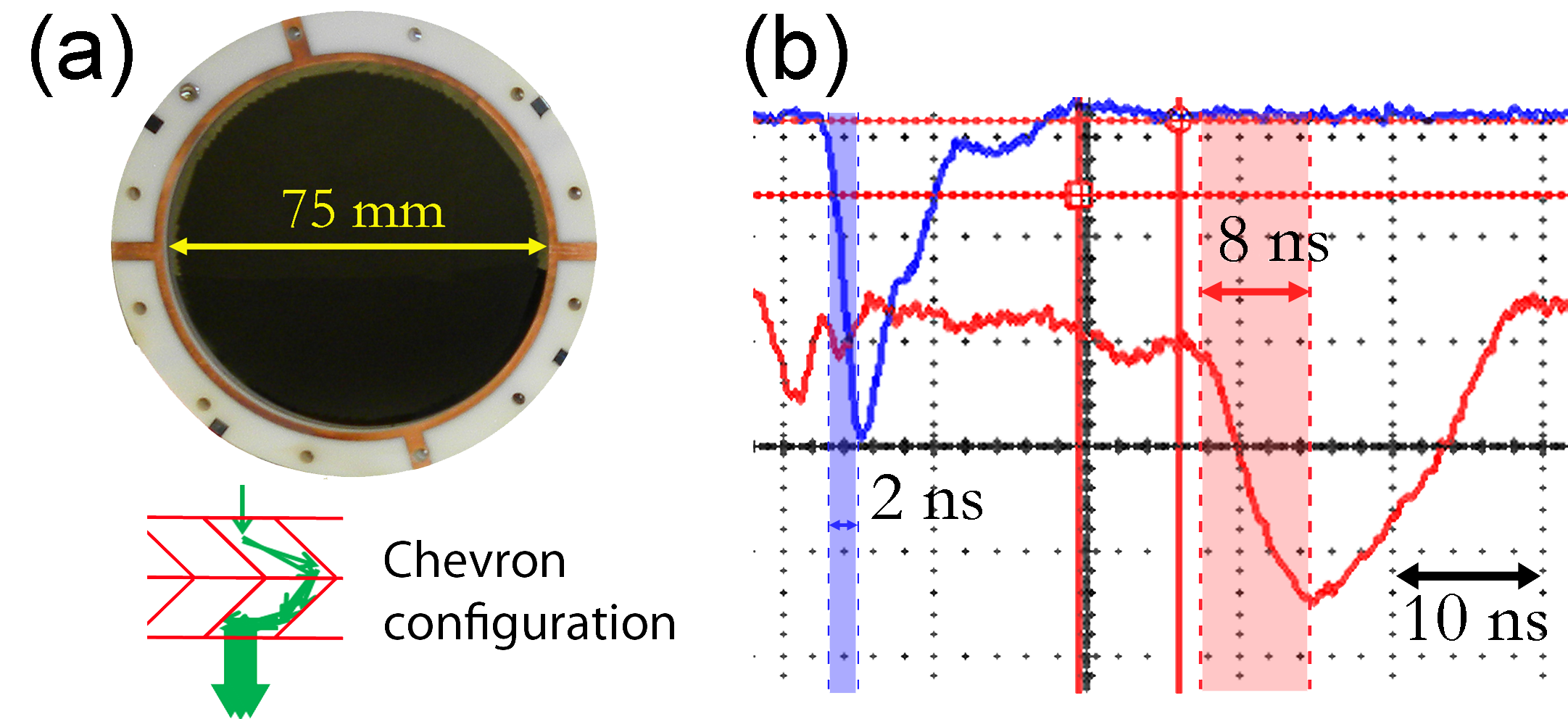}
\caption{\label{FIG:MCP2pan}(Color Online)(a) MCPs used in the present setup are 75 mm in diameter with 12 $\mu$m channel diameter.  Two plates are stacked in a ``Chevron configuration'' for optimal performance. (b) Inverted MCP signal (blue) has a rise time of approximately 2 ns (blue band). An accompanying DLA signal (red) has a rise time of approximately 8 ns (red band).}
\end{figure}
The MCPs are electron multipliers. Two impedance-matched MCPs are arranged in a ``Chevron configuration'' (see Fig.~\ref{FIG:MCP2pan}) for optimum performance, and stacked between two insulating ceramic rings that have been metalized in chosen locations for the purpose of applying the desired voltage.  The SE strike the top of the MCP which is grounded relative to the high positive potential applied to the base (2200 to 2500 V).  The signal produced by the MCP is fast having typical rise times of 2 ns and full width of 10 ns (see Fig.~\ref{FIG:MCP2pan}). This signal is used as a time stamp for charged particles passing through the conversion foils.  The highly multiplied electrons are ejected onto the DLA where they are collected to reconstruct the position of the interactions between the fission fragment and the conversion foil.  Typical DLA signals have 8 - 10 ns rise times and 25 ns full widths.  The DLA consists of two orthogonal dual-wire wrappings for $x$ and $y$ position determination. The constituents of the wire pairs are labeled $signal$ and $reference$ by convention and are completely isolated from each other when working properly. Wire pairs had a spacing of approximately 1 per mm.  The DLA floated at high positive voltage (2300 to 2600 V) employing a splitter with a +36 V battery on one side of the split to keep the $signal$ and $reference$ wires of the DLA at a small voltage difference.  The collected electrons propagate away from the spot where they struck the DLA in both directions producing two signals for each dimension (e.g., $x$$_{1}$ and $x$$_{2}$). By taking the time difference of the signals (e.g., $\Delta x$ and $\Delta y$), a two dimensional projection of the $x$ and $y$ position of the fission fragment can be interpreted.  

\subsection{\label{sec:eff} Detection Efficiency}
The efficiency of our time pick-off detectors will depend on several factors including the number of SE created, the energy of the electrons striking the MCP, the angle of incidence of electrons on the MCP, etc.  Maximum detection efficiency for a single electron is approximately 85\%. Each $\alpha$-particle or fission fragment will produce multiple SE with the SE yield depending the velocity and $Z$ of the particle \cite{Bethe37}. Tests of the detection efficiency of similar time pick-off detectors conducted at the University of New Mexico determined that this kind of time pick-off detector will be approximately 100\% (70\%) efficient for detecting fission fragments ($\alpha$-particles) \cite{UNMthesis}.

\subsection{\label{sec:DAQ}Signal Processing and Data Acquisition\protect}
In order to achieve the target temporal resolution, the %signal procession 
hardware needs to be optimized for fast timing signals. Thus, the data acquisition system must be designed to measure time differences on the order of 10s to 100s of picoseconds.  %Much care was taking in choosing 
The preamplifiers, discriminators and DAQ system that we used were chosen to be optimized for the fast timing requirements.

The five signals (one fast timing signal and four position signals) from each timing detector are amplified using ORTEC VT120C preamplifiers \cite{ortec}~developed for signals with fast rise times (e.g., microchannel plate signals). The signals from the detectors have rise times of approximately 2 ns; the amplifier has an intrinsic rise time of $<$1ns. The amplified signals are then sent through a Phillips Scientific 715 constant fraction discriminator (CFD) with 1 ns (5 ns) delay for timing (position) signals.  These CFDs are specified by the manufacturer to exhibit very small ($\pm$75ps) time walk.  A comparison of the time walk of several different discriminators for MCP signals made by Kosev in 2007 showed that among several CFDs, the Phillips Scientific CFD module had the best performance \cite{thesis}.

A VME based data acquisition from CAEN S.p.A \cite{caen}~is used for the digitization of signals. The NIM signals from the discriminators are sent to a time-to-digital converter (TDC) CAEN model V1290N.  Ten of 16 available channels were used to accommodate the signals from two timing detectors. The fast timing signal from the $stop$ detector is used as the external trigger of the TDC. The timing of other signals is analyzed relative to the $stop$ detector timing signal on an event-by-event basis. The TDC stored the time differences with 25 ps least significant bit, and has a resolution of 35 ps. The TDC is read out by a crate controller to a PC over an optical connection. The data is monitored and stored to disc using the MIDAS \cite{midas}~software package.

\section{\label{sec:ExperimentandDP}Detector Performance}

\begin{figure*}
\includegraphics[width=0.95\textwidth]{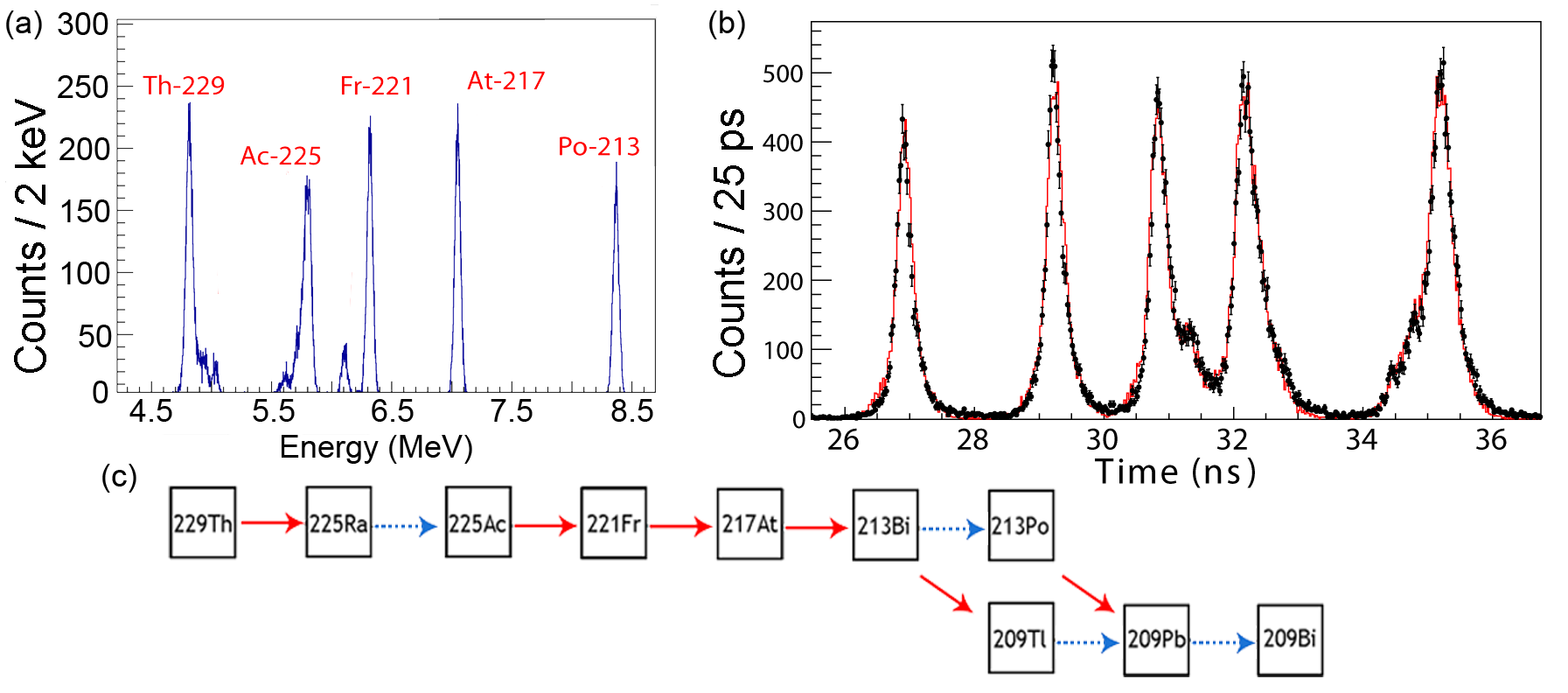}
\caption{\label{Fig:dkTH}(Color Online) (a) Simulated $\alpha$-particle energy spectrum for $^{229}$Th and its decay chain given realistic branching ratios.  (b) TOF spectrum for $^{229}$Th and its decay chain.  Simulation (red line) of 200 ps timing resolution reproduces the data (black dots).  Details of the simulation are discussed in section \protect{\ref{sec:sim}} (c) Decay scheme for $^{229}$Th and its decay chain. Solid red arrows indicate alpha decay.  Dotted blue arrows indicate beta decay.  }
\end{figure*}
Given the fixed distance between the $start$ and $stop$ time pick-off detectors, the timing of the fission fragments is generally predictable.  A fragment with 1 MeV per nucleon will spend approximately 50 ns traveling 70 cm. To achieve 0.5\% timing precision, the TOF measurement requires a precision of 250 ps. 
Well known $\alpha$-particle energies from $^{229}$Th and its decay chain and $^{252}$Cf were used as a standard to determine the timing and position properties of these detectors. 
    
The source was mounted a few cm from the first conversion foil.  The $\alpha$-particles with suitable trajectories pass through the time pick-off detector, traverse the length of the chamber, and pass through the second time pick-off detector.  The second time pick-off detector is a mirror image of the first time pick-off detector (see Fig.~\ref{Fig:TOFdiagram}).  Thus, the $\alpha$-particle encounters the mirror grids of the second mirror before traversing the second carbon foil.  SPIDER uses forward-biased SE from the first foil, and backward-biased SE from the second foil.  This has been shown to have high efficiency by Kosev \cite{thesis}~and maximizes the TOF distance because of the orientation of the time pick-off detector modules.

\subsection{\label{sec:TimRez}Timing Resolution}
The $\alpha$-particles from $^{229}$Th and its decay chain are ideal for testing the timing of the present system because the source has six fairly well separated energy groups (4.8, 5.8, 6.1, 6.3, 7.05, and 8.4 MeV) which result from five strong alpha emitters ($^{229}$Th, $^{225}$Ac, $^{221}$Fr, $^{217}$At, and $^{213}$Po respectively).  The decay daughter $^{213}$Bi is a weak alpha emitter which hardly contributes (see Fig.~\ref{Fig:dkTH}).  The first timing tests using $^{229}$Th and its decay chain were performed using small carbon foils suspended across rings with a 1-2 cm diameter aperture.  A 9.9 mm diameter collimator was placed near the center of the TOF chamber in order to restrict acceptance only to flight paths which were nearly centered (See Fig.~\ref{Fig:TOFdiagram}). This restriction makes the TOF spectrum indicative of the attainable timing precision without requiring position information. Figure \ref{Fig:dkTH} shows the results of these tests compared with simulations indicating 200 ps timing resolution. A broader timing component ($\sim$680 ps) was observed to  contribute to these first timing tests.  Later tests conducted with $^{252}$Cf confirmed that this broad timing component was removable with optimized discriminator settings. 

\begin{figure}
\includegraphics[width=0.45\textwidth]{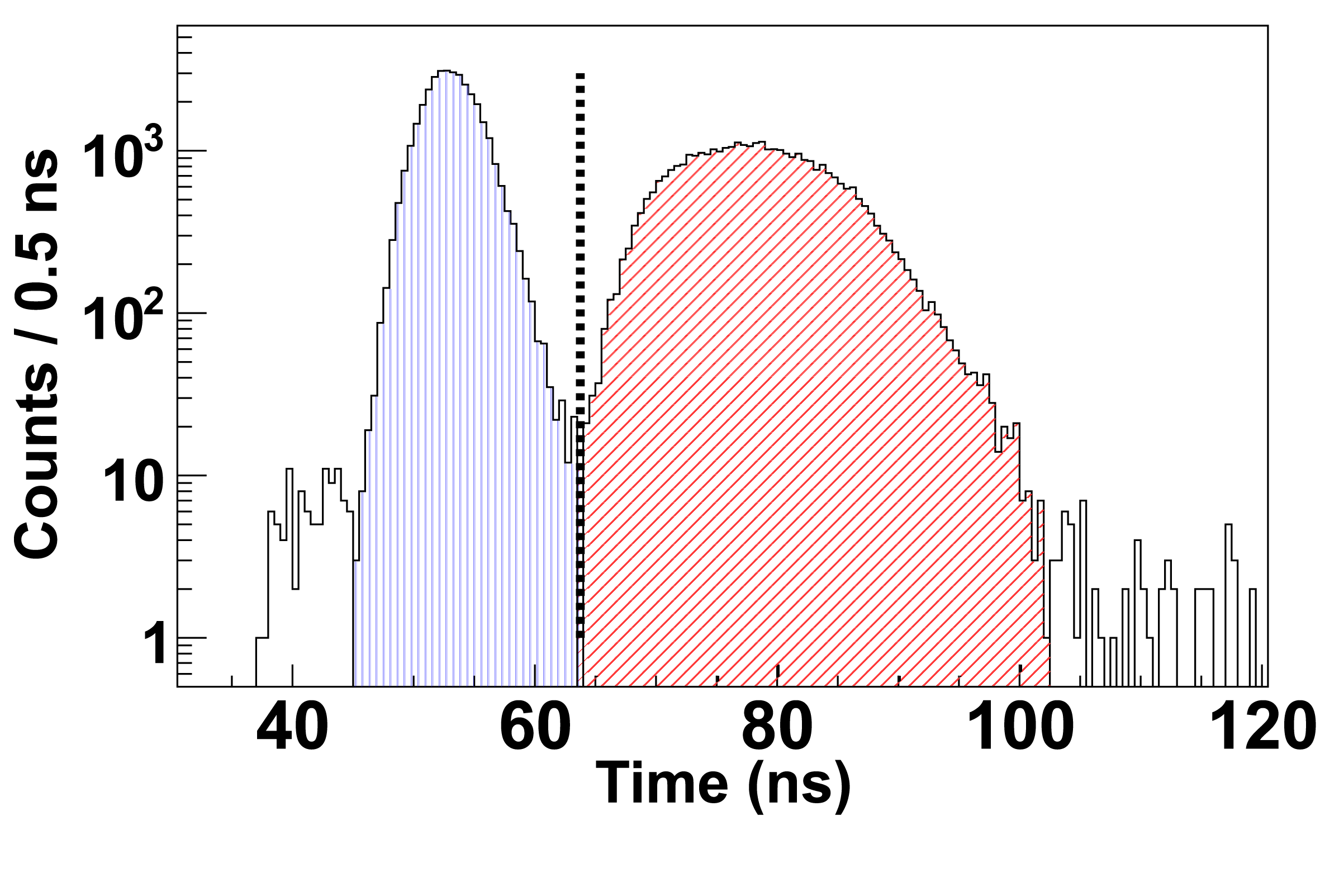}
\caption{\label{Fig:Cali}(Color Online) TOF spectrum for spontaneous fission fragments from $^{252}$Cf for time pick-off detectors separated by 700.0 mm.  The fast (blue) and slow (red) fission fragments have almost no overlap under these conditions and the number of integrated counts in each region are in agreement to within $<$ 1\%. The enhanced background at 40 ns is caused by the strong 6.217 MeV $\alpha$-particle from $^{252}$Cf. Most $\alpha$-particle events were below threshold.}
\end{figure} 
Californium-252 is an excellent source for testing the timing of our time pick-off detectors because it emits $\alpha$-particles and spontaneous fission fragments.  $^{252}$Cf emulates the experimental conditions to which we will ultimately subject our detectors. Tests with $^{252}$Cf were performed using conversion foils that were rectangular (8.7 cm $\times$ 6.7 cm) Mylar foils with a thin evaporated gold layer.  Fig. \ref{Fig:Cali} shows two very well separated regions corresponding to the light and heavy fission fragments from $^{252}$Cf. Signals from $\alpha$-particles were below the threshold settings in the spectrum shown in Fig. \ref{Fig:Cali}. The integral of the light fission fragment peak and the heavy fission fragment peak are identical to within 1\% indicating that the efficiency for detecting light and heavy fragments is essentially identical.   

\subsubsection{\label{sec:PosTes} Spatial Resolution}
Given the spacing of the wires on the DLAs, 1 mm spatial resolution or better should be possible.  The precision of the optics in our electrostatic mirror impose limits on the resolution of the image projected onto the MCPs. Understanding the position resolution of our system is important because we will reconstruct event-type data.  Though these corrections are small, they improve the overall precision of the velocity measurement.  

\begin{figure}
\includegraphics[width=0.45\textwidth]{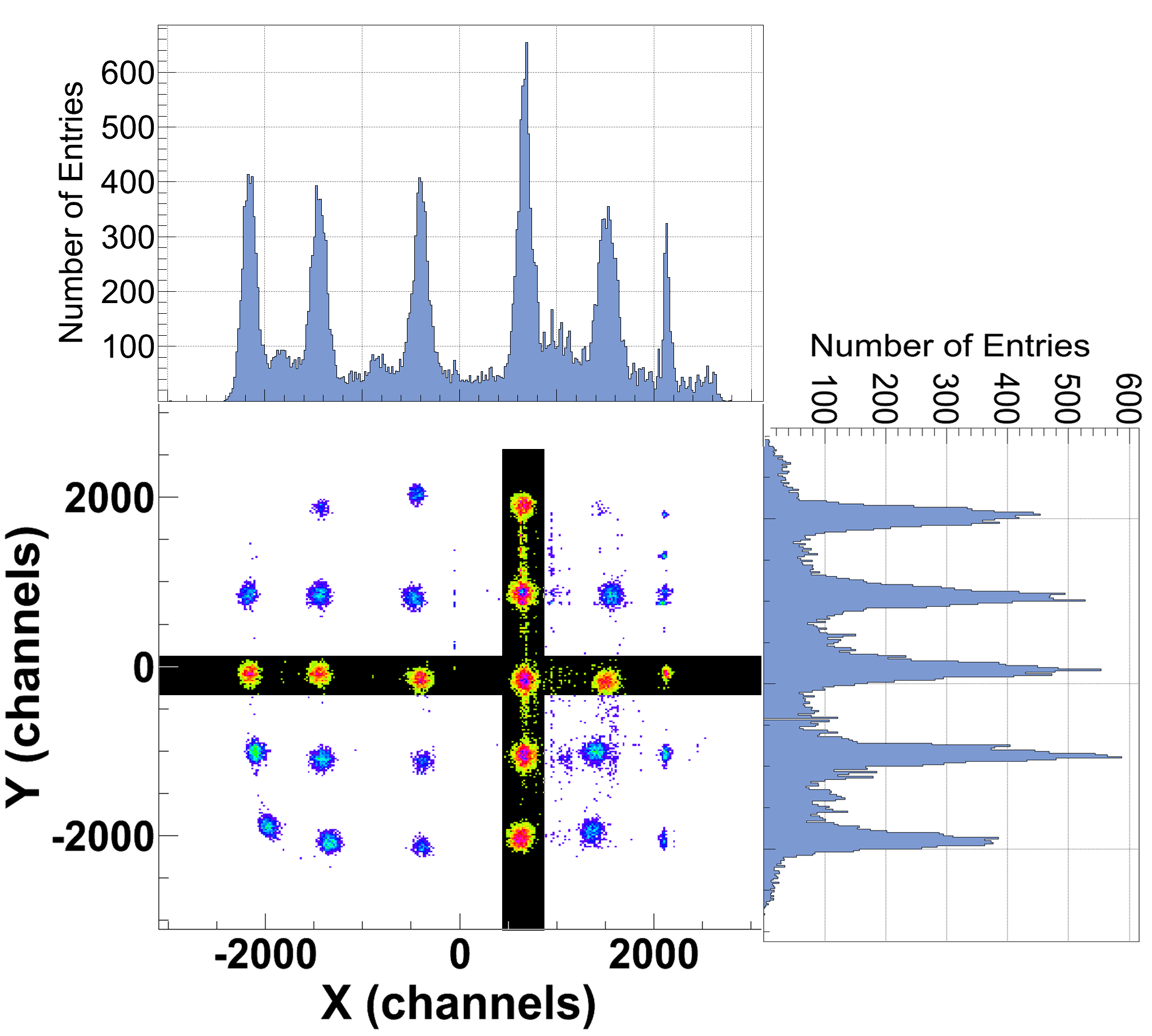}
\caption{\label{Fig:masksetup}(Color Online) The image produced when using a mask of 1 mm diameter holes arranged in a 1 cm $\times$ 1 cm matrix. A projection of a representative row and column indicated position resolution of approximately 2 mm FWHM. To remove signals resulting from dark current and show the mask image more clearly the 2D histogram is shown with a 10 count/bin threshold.  Corrections to warping of the image near the edges of the MCP have not yet been made. }
\end{figure}

A mask was constructed from 1 mm thick aluminum with 1 mm diameter holes arranged in a 1 $\times$ 1 cm matrix (see Fig.~\ref{Fig:masksetup}). We placed the mask between the source and the first foil as close to the foil as possible.
The mask tests revealed that the present system can resolve the location that a charged particle traversed the foil to within 2 mm.  For two conversion foils separated by 700.0 $\pm$ 0.5 mm, the uncertainty of the $x-y$ spatial position introduces less than 0.2 mm of additional uncertainty to the path length for an overall uncertainty of $\leq$ 0.1\%.
  
\section{\label{sec:sim}Simulations}
In an effort to understand better some of the individual contributions to the uncertainty in fragment velocity measurements, we drafted a simple simulation of the $\alpha$-particles from $^{229}$Th and its decay chain using \root~\cite{rootcode} and \srim~\cite{srimcode} software. The program simulates energy losses that result from straggling during the epoch of an $\alpha$-particle that goes through both detectors.  In addition to the energy losses and broadening that result from straggling, a user-defined broadening term is incorporated to adjust for the remaining sources of broadening (e.g., spread in secondary electron transit times, analog hardware, digitization).  %Thus, the energy loss and straggling of $\alpha$-particles through the source material and the first and second carbon foils is considered.  
Straggling out of the source material reduces and broadens the $\alpha$-particle energy initially, but has no real effect on the precision of the TOF measurement.  The straggling in the first conversion foil will reduce the energy of the $\alpha$-particle and broaden the attainable timing precision.  The second foil will have the same affect on the $\alpha$-particle, however, only the energy broadening will affect the TOF measurement.  

The \srim~(Stopping Range of Ions in Matter) software simulates the energy losses of ions traveling in and through matter.   A series of \srim~simulations were run for $\alpha$-particles of energies ranging from 4.5 to 8.5 MeV through carbon foils of thicknesses ranging from 20 to 100 \mg.  The data sought from these simulations were the average energy loss and the width of the distribution of the energy loss of the $\alpha$-particles. Histograms of energy losses for 10000 $\alpha$-particles  were fit with Gaussian functions.  The mean and sigma of the Gaussian fit are measures of the average energy loss and energy spread for $\alpha$-particles.   Compilations of the parameters obtained from these simulations over the ranges mentioned above were fit, often with a simple linear function.  Under the assumption of a good fit to the simulations, the average energy loss and width of the distribution can be interpolated for $\alpha$-particles  and foil thicknesses within the simulated ranges.  A separate \srim~simulation of the straggling of $\alpha$-particles out of the source material was generated making some assumptions about the source thickness and other properties of the source material.  

Next, the formulas derived from \srim~simulations are imported into a \root~simulation. The \root~simulation chooses an $\alpha$-particle from $^{229}$Th and its decay chain and subjects it to the straggling effects listed above.  That is, in the \root~simulation a number is randomly sampled from a Gaussian function representing the correct energy loss distribution for the given $\alpha$-particle energy and foil thickness.  After straggling out of the target and through the first foil, the energy and velocity of the simulated $\alpha$-particle is known.  The distance between the two foils is an input parameter which determines the time that the $\alpha$-particle spends drifting between the foils.  After straggling through the second foil, the time difference is recorded.  

Figure \ref{Fig:dkTH} shows the comparison of a simulated $\alpha$-particles and data.  A simulation with two timing components (narrow and wide) was found to best reproduce the data.  The proportion of signals exhibiting the narrow timing distribution as compared to the wide timing distribution was approximately 2:1 with the narrow component having 200-ps FWHM timing and the wide component having 680-ps FWHM timing.  In subsequent tests with a $^{252}$Cf source, the presence of a wide component on the $\alpha$-particle peak was absent.  It is believed that optimizing the discriminator settings was the primary contributor to the removal of the wide timing component.

%With the possible flight paths of the $\alpha$-particles constrained by the 50 - 70 cm distance between carbon foils and 

%\begin{eqnarray}
% \epsilon \propto d\Omega \approx~4\pi~~~~~~~~~~~~~~~~~~~~~~~~~~~~~~~~~~~~~~~~~~\nonumber\\
%  - 2\pi\times\left[\left(1-\frac{L/2 - z}{\sqrt{(L/2-z)^{2} + r^{2}}}\right)\right.~~~~~~~~~~~~\nonumber\\
%  ~~~~~~~~~~~\left.+\left(1-\frac{L/2 + z}{\sqrt{(L/2+z)^{2} + r^{2}}}\right)\right]
%\end{eqnarray}

\section{\label{sec:results}Summary and Conclusions\protect}
The goal of this work was to optimize and characterize the timing properties of position-sensitive, time pick-off detectors to be used in SPIDER. In order to measure the mass of light fission fragments to a single mass unit, the detectors must exhibit timing resolution of better than 250 ps (FWHM).  The timing resolution of the time pick-off detectors presented here was verified to be at least 200 ps (FWHM) using $\alpha$-particles from $^{229}$Th and its decay chain and $^{252}$Cf.  Optimal timing measurements were realized when (1) the distance between time pick-off detectors was maximized (70 cm in the present arrangement), and (2) carbon foils of 20 \mg~thickness were used.  Mounting thinner carbon foils was not attempted.  With 200 ps (FWHM) timing resolution for fission fragments, coupled with $E_{k}$ measurements with 400 keV energy resolution, SPIDER will successfully resolve light fission fragment masses to better than 1 mass unit.

\section*{Acknowledgement}
This work was performed under the auspices of the U.S. Department of Energy at Los Alamos National Laboratory operated by the Los Alamos National Security, LLC under Contract No. DE-AC52-06NA25396.% for LDRD project 20120077DR.

\bibliographystyle{elsarticle-num}
\bibliography{SPIDER_InstrumNIM}% Produces the bibliography via BibTeX.

\end{document}